\documentclass[twocolumn]{webofc}

\usepackage[varg]{txfonts}   
\usepackage{hyperref}
\usepackage{url}
\usepackage{subcaption}
\hypersetup{colorlinks=true,citecolor=blue,urlcolor=blue,linkcolor=blue}
\begin{document}
%
\title{Status of Barrel Imaging Calorimeter in Korea for the Electron-Ion Collider}
%
%

\author{\firstname{Jeongsu} \lastname{Bok}\inst{1}\fnsep\thanks{\email{jeongsu.bok@cern.ch}}
}

\institute{Department of Physics, Pusan National University, Busan, South Korea}

\abstract{
The Electron-Ion Collider (EIC) is a next-generation particle accelerator facility designed to probe the fundamental structure of matter such as the origins of nucleon mass, spin, and the dynamic behavior of quarks and gluons within nucleon and nucleus. As the electromagnetic calorimeter in the barrel region, the Barrel Imaging Calorimeter (BIC) is tasked with precise energy measurements of electrons and photons as well as efficient separation of these particles from background pions. The BIC integrates Pb/SciFi sampling layers and AstroPix silicon pixel sensors for three-dimensional shower imaging. The Korean group has actively contributed through silicon chip testing, module assembly, prototype development, beam test, readout system design, and detailed simulations. This presentation highlights the recent progress and plans for the R\&D of the Barrel Imaging Calorimeter in Korea.
}
\maketitle
\section{Introduction}

The Electron-Ion Collider (EIC)~\cite{AbdulKhalek:2021gbh} will explore how quarks and gluons build the mass, spin, and structure of visible matter. It will operate with polarized electron and hadron beams (5--18~GeV electrons and 41--275~GeV protons or ions), reaching luminosities up to $10^{33}$--$10^{34}$~cm$^{-2}$s$^{-1}$. Precision 3D imaging of nucleons and nuclei requires high-performance calorimetry in the barrel region of the general-purpose EIC detector. The ePIC (The Electron-Proton/Ion Collider) experiment is the first experiment at EIC.

The Barrel Imaging Calorimeter (BIC) is designed as the electromagnetic calorimeter in the barrel region $-1.7<\eta<1.4$, where $\eta = -\ln[\tan(\theta/2)]$ and $\theta$ is the polar angle with respect to the beam axis. The energy resolution of $10\%/\sqrt{E}\oplus(2-3)\%$ and electron--pion separation of $10^4$ at low momenta are required~\cite{AbdulKhalek:2021gbh}. Additionally, it should provide good photon reconstruction down to $\sim$100~MeV and $\pi^0$ discrimination up to $\sim$10~GeV. These requirements motivate an imaging-type electromagnetic calorimeter combining Lead-Scintillating Fiber (Pb/SciFi) layers and silicon pixel layers.

\section{Design Concept of the BIC}
\begin{figure}[h]
    \centering
    \includegraphics[width=\linewidth]{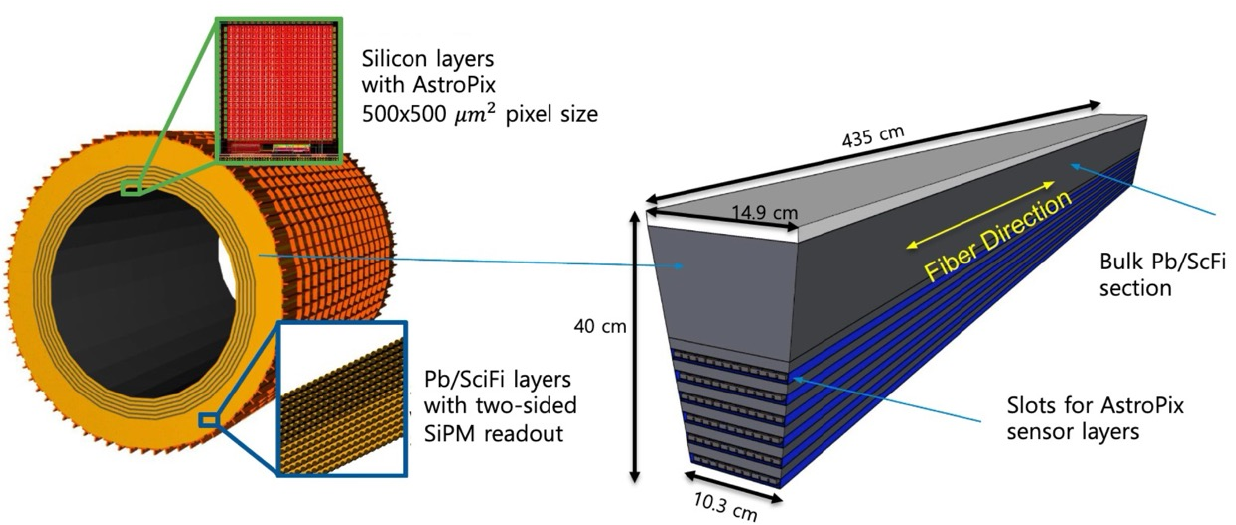}
    \caption{Design of the Barrel Imaging Calorimeter.}
    \label{fig:BICdesign}
\end{figure}

The BIC combines two subsystems: (i) Pb/SciFi sampling layers for energy measurement and (ii) AstroPix silicon pixel layers for shower imaging as shown in Figure~\ref{fig:BICdesign}. 
The Pb/SciFi sampling calorimeter is based on the BCAL at GlueX experiment and the AstroPix silicon pixel detector was developed for the NASA Amego-X mission~\cite{GlueX:2020idb, Steinhebel:2025pra}.
The baseline design features 4 imaging layers of silicon pixel layers interleaved with 5 Pb/SciFi layers, followed by a bulk section of Pb/SciFi. The total radiation length is about 17~$X_0$, where $X_0$ is the characteristic length over which a high-energy electron loses all but $1/e$ of its energy by bremsstrahlung.
The sampling fraction of the calorimeter is approximately 10\%.
AstroPix layers provide fine position resolution and 3D shower profiling, while the sampling calorimeter determines energy. The full barrel calorimeter will cover about 100~m$^2$ using $\sim$2.5$\times$10$^5$ AstroPix chips ($\sim$2$\times$2~cm$^2$ each).

\section{Korean Contributions and Prototype Development}
In the BIC project, the Korean group is contributing to several core tasks: (i) AstroPix chip testing and wafer-level evaluation, (ii) Pb/SciFi production, (iii) Readout box development, (iv) GEANT4 simulation and system-level performance studies.

\begin{figure}[h]
\begin{center}
\begin{subfigure}{0.45\linewidth}
\includegraphics[width=0.9\linewidth]{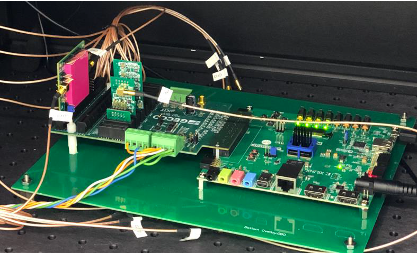} 
\label{fig:subim1}
\end{subfigure}
\begin{subfigure}{0.45\linewidth}
\includegraphics[width=0.9\linewidth]{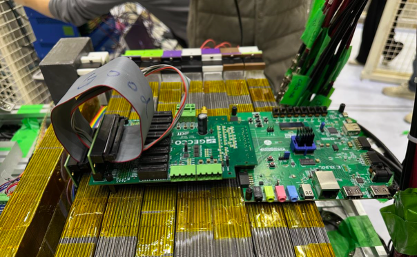}
\label{fig:subim2}
\end{subfigure}

\caption{AstroPix v3 single chip setup for lab test (Left) and flexible cable test at KEK PF-AR test beam line in March 2025 (Right).}
\label{fig:astro}
\end{center}
\end{figure}

The Korean group is preparing the mass chip test. For the mass chip test, multi-line single-chip test system is in consideration because one chip handler can cover multiple test stations depending on the test procedure and time. During the recent beam test at KEK in March 2025, connection with a flexible cable has been tested as shown in Fig.~\ref{fig:astro}. The flexible cable would be also useful for the future beam test in order to locate AstroPix modules between Pb/SciFi layers.

\subsection{Pb/SciFi Prototype Modules}
\begin{figure}[h]
    \centering
    \includegraphics[width=\linewidth]{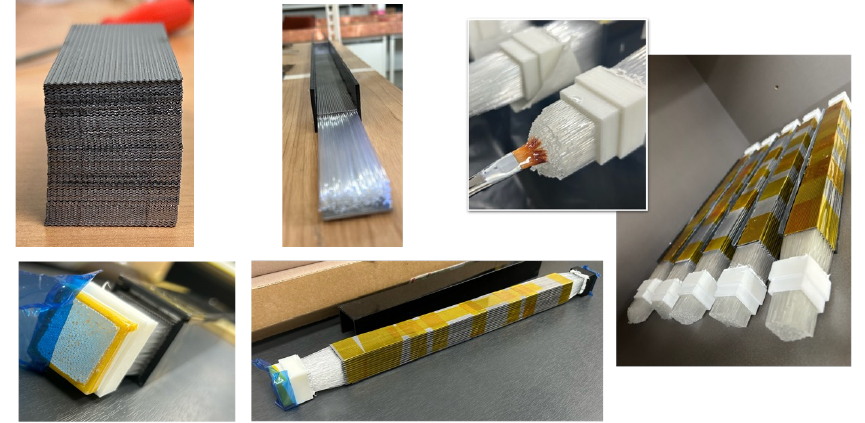}
    \caption{Production procedure of a unit module includes stacking of lead plates and scintillating fibers, curing, and polishing.}
    \label{fig:moduleproduction}
\end{figure}

Prior to the beam tests, Korean group prepared a set of prototype modules. The dimension of a unit module is $\rm 32\times3\times3\ cm^3$. To produce a unit module, Pb plates with 0.5~mm thickness are shaped with press, then stacked with scintillating fibers (1~mm diameter), cured, and polished as shown in Fig.~\ref{fig:moduleproduction}. A total of 33 modules were produced in Korea, including one with a light-guide prototype. The modules were instrumented with PMTs or SiPMs in the beam tests. Bundling and connection between the fibers and the PMT (or SiPM) were made using 3d-printed cases.

\section{Beam Test Results}
\subsection{CERN PS T10 (August 2024)}
\begin{figure}[h]
    \centering
    \includegraphics[width=\linewidth]{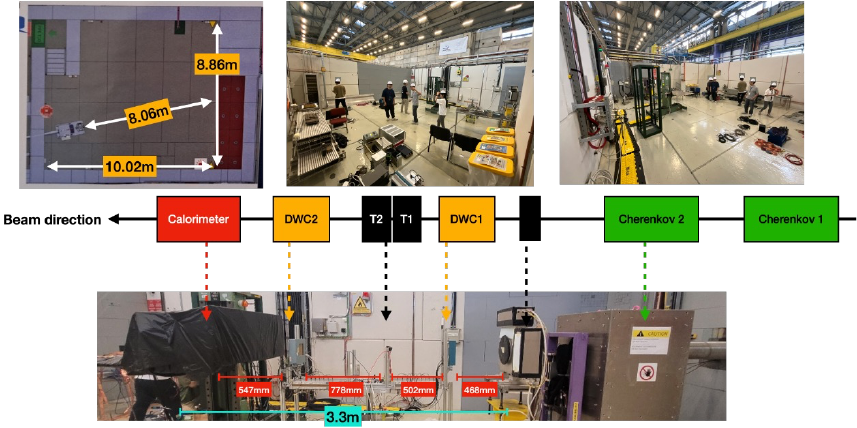}
    \caption{Experimental setup in the beam test at CERN PS T10.}
    \label{fig:CERN2408_setup}
\end{figure}

\begin{figure}[h]
    \centering
    \includegraphics[width=\linewidth]{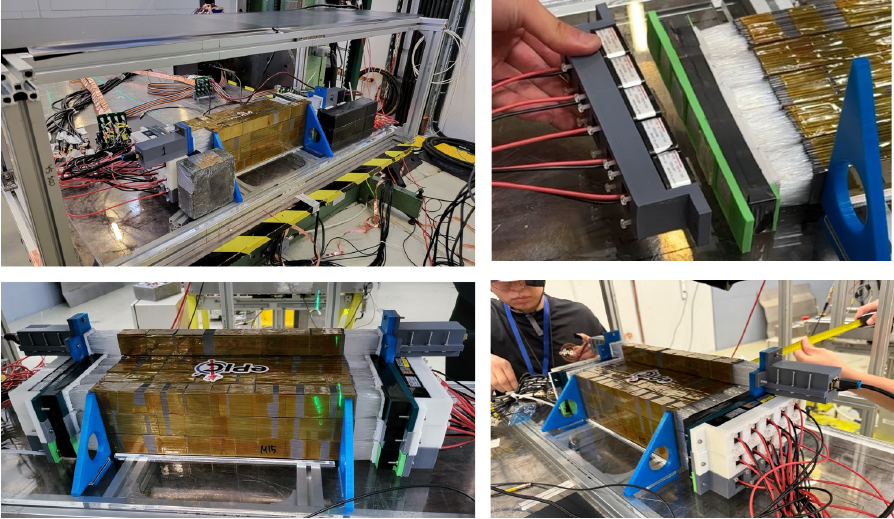}
    \caption{The total dimension of the Pb/SciFi was $\rm 32(h)\times 9(w)\times 15(d)\ cm^3$. A 3$\times$5 array of unit modules was formed. A Pb/SciFi unit module equipped with SiPM was installed on top of it.}
    \label{fig:CERN2408_assembly}
\end{figure}

\begin{figure}[h]
    \centering
    \includegraphics[width=0.8\linewidth]{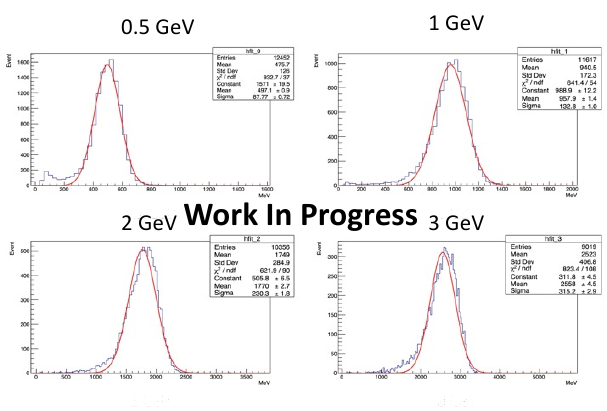}
    \caption{Response of the Pb/SciFi from 0.5, 1, 2, and 3 GeV/$c$ electron beam in the beam test at CERN PS T10.}
    \label{fig:CERNresult}
\end{figure}

The first beam test was conducted using 0.5--3~GeV/$c$ electron beams at CERN PS T10 in 2024. The beam was a mixture of charged pions, electrons, and other particles. The identification of electron was performed using the Cherenkov counters at the beam line. As shown in Fig.~\ref{fig:CERN2408_setup}, a trigger composed of two 1-cm-thick scintillators and PMTs was installed. Two Delay Wire Chambers were used to reject scattered beam particles. The calorimeter was composed of 3$\times$5 array of unit modules, giving a total detector size of $\rm 32(h)\times 9(w)\times 15(d)\ cm^3$ as shown in Fig.~\ref{fig:CERN2408_assembly}. The depth corresponds to $\sim$11 $\rm X_0$. The signal from PMTs was collected by Flash ADC. During the experiment, we conducted electron calibration successfully and collected data for 0.5--3 GeV/$c$ electrons. Data are being analyzed for obtaining energy resolution as shown in Fig.~\ref{fig:CERNresult}.

\subsection{KEK PF-AR (March 2025)}
\begin{figure}[h]
    \centering
    \includegraphics[width=\linewidth]{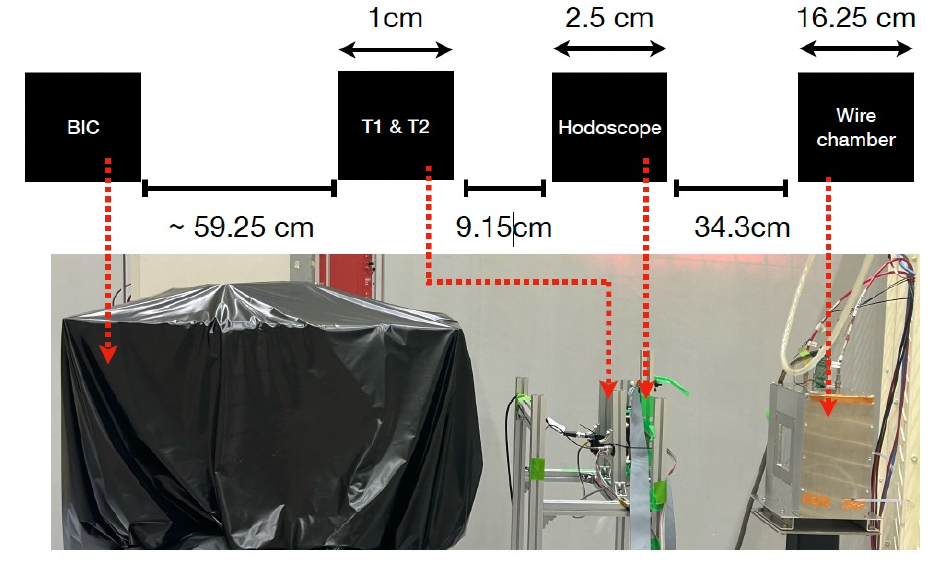}
    \caption{Beam test setup at KEK PF-AR testbeam line. Two trigger counters and a hodoscope with scintillating fibers were installed in front of Pb/SciFi calorimeter.}
    \label{fig:KEKsetup}
\end{figure}

\begin{figure}[h]
    \centering
    \includegraphics[width=0.8\linewidth]{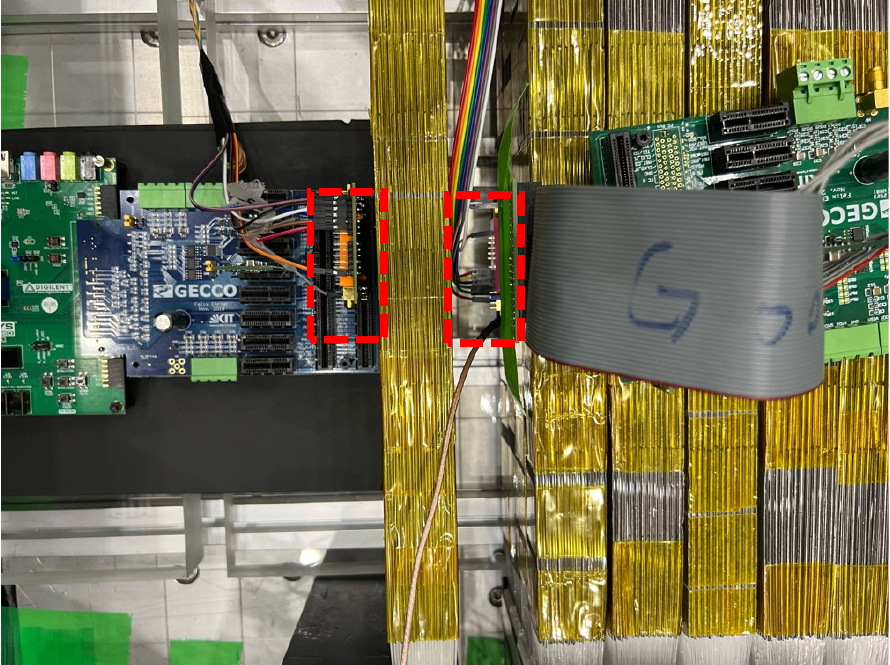}
    \caption{During the beam test at KEK PF-AR, AstroPix v3 was located between Pb/SciFi layers using an extension cable. In front of the Pb/SciFi layer, a single-chip setup with GECCO board was installed.}
    \label{fig:KEKastro}
\end{figure}

\begin{figure}[h]
    \centering
    \includegraphics[width=\linewidth]{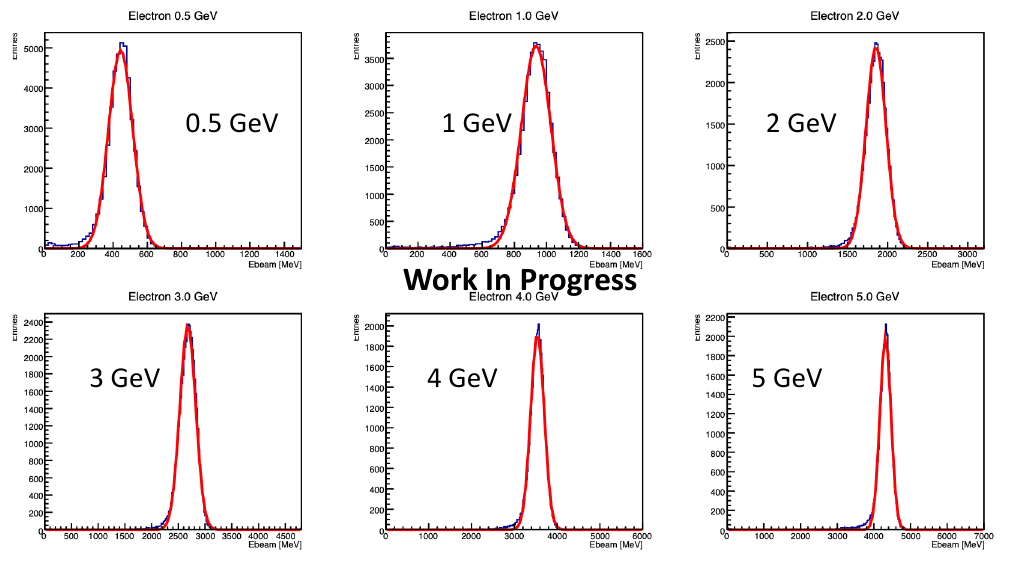}
    \caption{Response of the Pb/SciFi from 0.5 to 5 GeV/$c$ electron beam in the beam test at KEK PF-AR.}
    \label{fig:KEKresult}
\end{figure}

The second beam test used a 4$\times$7 array of Pb/SciFi modules ($\sim$15~X$_0$ depth) and included AstroPix layers placed between Pb/SciFi modules as shown in Figs.~\ref{fig:KEKsetup} and \ref{fig:KEKastro}. Electron beams of 0.5--5~GeV were used to study the energy and position response in Fig.~\ref{fig:KEKresult}, longitudinal and transverse profiles.


In the beam test at KEK PF-AR, AstroPix signals recorded through the GECCO + FPGA development board were recorded simultaneously with signals from other detectors including Pb/SciFi using the new DAQ system produced in Korea. Preliminary correlations between beam positions on AstroPix and wire-chamber drift times were checked. This demonstrates the feasibility of synchronized data taking of silicon layers and Pb/SciFi layers, a key milestone for 3D shower imaging. This will be very useful for performance test under beam environment.

\section{Summary and Outlook}
The Barrel Imaging Calorimeter is a core detector for the ePIC experiment in EIC. The Korean group is making significant contributions in silicon testing, Pb/SciFi module production, readout box, and system testing. Especially Korean group performed beam tests at CERN PS and KEK PF-AR recently. We successfully collected electron-beam data, and the analysis is ongoing. In addition, integrated DAQ test for AstroPix and Pb/SciFi under beam environment represents a step toward 3D shower imaging which is an essential capability of the BIC. Further beam tests in 2025 and 2026 will be critical to validate the system performance and the preparation of larger prototypes will be a guide for the mass production. Upcoming beam tests at KEK and CERN will combine AstroPix and Pb/SciFi modules for 3D imaging of electromagnetic showers. Also, 70~cm-long bulk sector prototype is under preparation, including readout with HGCROC and SiPM. 

\end{document}